\documentclass[sigplan,10pt,screen]{acmart}

\AtBeginDocument{%
  \providecommand\BibTeX{{%
    Bib\TeX}}}

\setcopyright{acmlicensed}
\acmDOI{10.1145/3759429.3762626}
\acmYear{2025}
\copyrightyear{2025}
\acmISBN{979-8-4007-2151-9/25/10}
\acmConference[Onward! '25]{Proceedings of the 2025 ACM SIGPLAN International Symposium on New Ideas, New Paradigms, and Reflections on Programming and Software}{October 12--18, 2025}{Singapore, Singapore}
\acmBooktitle{Proceedings of the 2025 ACM SIGPLAN International Symposium on New Ideas, New Paradigms, and Reflections on Programming and Software (Onward! '25), October 12--18, 2025, Singapore, Singapore}
\received{2025-04-25}
\received[accepted]{2025-08-11}


\usepackage[frozencache,cachedir=.]{minted2}
\usepackage{prettyref}
\newrefformat{sec}{Section~\ref{#1}}
\newrefformat{alg}{Algorithm~\ref{#1}}
\newrefformat{fig}{Figure~\ref{#1}}
\newrefformat{tab}{Table~\ref{#1}}
\newrefformat{def}{Definition~\ref{#1}}
\newrefformat{app}{Appendix~\ref{#1}}
\newrefformat{cor}{Corollary~\ref{#1}}
\newrefformat{li}{Line~\ref{#1}}
\newcommand\pref[1]{\prettyref{#1}}

\newcommand\trex{\textsc{TReX}}
\usepackage{graphicx}

\usepackage{listings}
\lstdefinestyle{myLuastyle}
{
    language         = {[5.0]Lua},
    basicstyle       = \ttfamily,
    showstringspaces = false,
    upquote          = true,
}
\usepackage[noend,ruled,linesnumbered]{algorithm2e}
\usepackage{subcaption}
\usepackage{tikz}
\tikzset{ gnode/.style={draw,circle}}
\tikzset{gcnode/.style={draw,rectangle}}

\usepackage{pgf}
\usepackage{pgfplots}
\usepgfplotslibrary{external}
\setminted{fontsize=\small}

\usepackage{pdfpages}

\begin{document}

\title{Literate Tracing}

\author{Matthew Sotoudeh}
\orcid{0000-0003-2060-1009}
\affiliation{%
  \institution{Stanford University}
  \city{Stanford}
  \country{USA}
}
\email{sotoudeh@cs.stanford.edu}


\begin{abstract}
    As computer systems grow ever larger and more complex, a crucial task in
    software development is for one person (the system expert) to communicate
    to another (the system novice) how a certain program works.
    This paper reports on the author's experiences with a paradigm for program
    documentation that we call \emph{literate tracing}.
    A literate trace explains a software system using annotated, concrete
    execution traces of the system.
    Literate traces complement both in-code comments (which often lack global
    context) and out-of-band design docs (which often lack a concrete
    connection to the code).
    We also describe \trex{}, our tool for making literate traces that are
    interactive, visual, and guaranteed by construction to be faithful to the
    program semantics.
    We have used \trex{} to write literate traces explaining components of
    large systems software including the Linux kernel, Git source control
    system, and GCC compiler.
    %
\end{abstract}




\maketitle

\section{Introduction}
The dream of production-quality computing systems small enough to be casually
studied and customized is dead.
Our kernels, compilers, and browser engines have grown large and unwieldy.
It often seems impossible for any one person to understand whole systems;
even experienced programmers might need days or weeks to understand a subsystem
well enough to make a seemingly trivial change.
%

Users rightfully demand that their software support an ever-growing set of
hardware and software configurations while being secure, reliable and
performant.
Despite virtuous attempts to reduce the size and complexity of modern software
systems~\cite{steps,cleancode,aposd}, these modern user demands fundamentally
conflict with small, hackable systems design.

Hence, we must design techniques that make it easier to bring novices
up-to-speed to understand and work with these complex systems.
Comments, for example, help give the new reader of a piece of code higher level
intuition for its behavior.
Unfortunately, because they are located in the code itself, they tend to be
most useful for \emph{local} context: it is hard for a comment in one module to
explain that module's relation to any number of possible interacting modules.
Furthermore, comments can obscure the code and cannot support explanatory
diagrams and long prose.

%
Documentation written separate from the source code can more easily focus on
the global structure of the code without being tied to any particular line in
the code, and encourages custom diagrams and extended prose.
Unfortunately, it tends to be \emph{too} high level, not giving the reader much
indication of how and where exactly the code implements this high-level design,
and silently falls out-of-sync as the codebase evolves.

This paper describes \emph{literate tracing}, a technique for explaining
software systems that helps bridge comments and out-of-band documentation.
\textbf{A literate trace is a document explaining how a software system works
by walking the reader through concrete execution traces of the system.}
Like out-of-band documentation, a literate trace is separate from the code.
What makes a literate trace special is that its structure is centered around
concrete execution traces of the program, helping the reader see exactly where
in the code a certain high-level action happens.
With proper tools~(\pref{sec:Tool}), authors can extract engaging
visualizations of a running program's internal state even for modern systems
software like the Linux kernel, GCC compiler, and Git source control system.
Portions of the trace that are relatively unimportant or uninteresting
can be skipped over without fear, as the reader can see any relevant effects of
the skipped-over portions on the concrete state.

From the documentation author's perspective, literate tracing is a rewarding
way to understand and document systems.
It involves reifying one's understanding of a system into a concrete object (the
literate trace) that can be shared with others.
This turns the task of \emph{reading} a program into the task of \emph{writing}
a program (the program that generates the literate trace and visualizations),
which is a task programmers often prefer.
This gives us hope for a future where writing and sharing a literate trace
becomes the natural next step after studying a large system.
Computer users everywhere could benefit from these traces, as they make the
software running on their computers easier to understand and modify.


\section{Literate Tracing by Example}
\label{sec:Examples}
The following pages contain an excerpt from our literate trace of Knuth's
program to compute the most frequent words in a file~\cite{mfw,litprog},
originally written as an early public examples of his literate programming
technique, which had the goal of inverting the relationship between comments
and source code.
\textbf{We now suggest the reader finish reading the excerpt in full before
returning to this analysis.}
Readers with additional time might appreciate also reading Knuth's literate
programming version for comparison~\cite{mfw} and our other example traces at
\url{https://lair.masot.net/trex}.

The program is not trivial, and no explanation would make it so.
%
%
But the literate tracing approach has the benefit of explaining the complicated
hash trie insertion routine via concrete examples and ample visualization of
the state of the internal data structure.
In fact, the need for a visualization of this data structure was one focus of
Doug McIlroy's critique of Knuth's literate program~\cite{lpcritique}.
Our literate tracing tool, \trex{}, makes it easy to generate such
visualizations programmatically, at document build time, from a running
instance of the program using the GDB API~(\pref{sec:Tool}).
This means the reader always sees a visualization of program state that is
guaranteed to reflect the actual source code.
Literate traces can display the same information at different levels, e.g.,
both the graph visualization and the underlying arrays encoding the trie in
memory, to help the reader understand how the high-level visualizations connect
to the code.

Literate tracing also departs from literate programming in that a literate
trace need not show the reader the entirety of the codebase, only the portions
that the author feels are most relevant.
Our literate trace skips over the entirety of the input/output code in favor of
focusing on the meaty hash trie structure; Knuth's literate program (because it
is itself the \emph{program}) must describe every line of code (including
input--output boilerplate and a ``somewhat tedious set of assignments'')
somewhere in the document.
In fact, literate traces are not even tied to explaining a single program: in
this very example, we found it useful to switch partway through to tracing a
version with a particularly bad pseudorandom number generator in order to
demonstrate the hash collision behavior.
Switching between multiple different program variants was also particularly
useful in our trace of Knuth's prime number generator program~\cite{litprog},
building it up one optimization at a time from a na\"ive implementation.

\subsection{HTML Literate Traces}
In addition to static literate traces written in \LaTeX{}, \trex{} supports the
generation of interactive literate traces in HTML.
The main benefit of the HTML version is that it supports easy creation of
\emph{interactive single-steppers} that the user can scrub through to see in
step-by-step detail the operation of the program.
In the interactive HTML version of this trace (available at
\url{https://lair.masot.net/trex}), the visualization is updated every time the
user steps to a new line, so they can see how the code affects the trie.
Another benefit is that HTML documents do not have page breaks, so it avoids
the annoyance common in PDF documents of worrying about figures, code, and
prose that cross page boundaries.
Finally, it frees the author from deciding exactly which steps to display to
the user and instead allows them to focus on the bigger picture visualizations
and structure of the document.

These benefits do come with tradeoffs.
Heavy reliance on interactive single-steppers robs the reader of the ability to
print and study traces offline.
And the existence of the single-stepper option tempts the author into trading
off prose explanations for longer and more fine-grained single-stepping traces
that the reader is forced to step through themselves.
This makes the author's life easier, but gives the reader much less direction
regarding what to focus on.

\subsection{Other Modes of Literate Tracing}
Many systems are far too large to expect any one reasonably sized document to
explain every part of the program.
We now discuss a few styles of literate tracing that are more applicable to
such scenarios.

\subsubsection{Explaining One Slice of the Program}
It is often useful to focus on one specific task that the system performs
rather than overwhelm the reader with a complete catalogue of its possible
behavior.
Understanding one slice of the program can allow the reader to make a specific
change that they are interested in, and the knowledge they learn from that one
slice frequently generalizes to other parts of the system as well.
For example, our trace of the Git source control system~\citep{git} works
through examples of how the \texttt{add} and \texttt{commit} commands work;
enough to get a sense of the internal structure of Git without getting bogged
down in the minutiae of every single Git command and feature.
The main benefit of literate tracing here is to break through abstraction
barriers: the reader sees exactly where in the code actions, like a file being
written to disk, occur in a way that is obscured by different layers of
abstraction (such as ref transactions and file locks) in the original code.

\subsubsection{Explain the Top Abstraction Level}
Other times, it is more profitable to take a \emph{shallow and wide}
approach, where all of the important program tasks are discussed, but details
of the implementation of some of those tasks are hidden.
\emph{(Continued \hyperlink{page.13}{after example}.)}

\clearpage



# 1 Most Frequent Words

Knuth's article *A Literate Program* describes a program to solve the following problem, posed by Jon Bentley:

> Given a text file and an integer $k$, print the $k$ most common words in the file (and the number of their occurrences) in decreasing frequency.

Knuth's original solution to this problem was written in Pascal via the WEB system of literate programming. This writeup contains a trace of a version of his code that we have manually translated to C. Code snippets and in-code comments in this trace are translated to C and adapted from the original Pascal literate programming version written by Donald Knuth (see Jon Bentley, Don Knuth, and Doug McIlroy. 1986. Programming pearls: a literate program. Commun. ACM 29, 6 (June 1986), 471–483. `https://doi.org/10.1145/5948.315654`).

## 1.1 Outline of the Program

Let's start by looking at the input and output of the program. The first line in the input file should be a number, indicating the value $k$ of how many words to print. The program then reads all of the remaining words in the document and prints the top $k$ in decreasing frequency, as requested.

```
2
xyz
xa
xa
```
The input file.

```
The 2 most common words, and their frequencies:
 xa 2
 xyz 1
```
The output file.

At a high level, the actual operation of the program is quite simple. There are five basic steps:

```
...
254:    get_word();
255:    while (WORD_LENGTH != 0) {
256:        p = find_buffer();
257:        if (p == 0)
258:            WORD_MISSED = 1;
259:        else if (COUNT[p] < MAX_COUNT)
260:            COUNT[p]++;
261:        get_word();
262:    }
...
265:    trie_sort();
...
268:    if (TOTAL_WORDS == 0) {
269:        printf("There are no words in the input!\n");
270:    } else {
...
278:        print_common(MAX_WORDS_TO_PRINT);
...
```

1. Read the words one-by-one from the file

2. Insert each one into a trie (**find_buffer**)

3. Keep a count for each word in the trie

4. Traverse the trie, sorting all of the words by their count

5. Print the first $k$ words from the sorted list

Experienced programmers will find little surprise in the routines for reading the input list of words and writing the output of most common words; in fact, our C version has replaced much of the manual string processing with **printf** and **scanf**. Instead, this trace will focus on the most interesting part of the program: Knuth's *hash trie* structure for storing a set of words.

# 2 The Hash Trie

Recall that a basic trie is a tree where the root represents the empty string `""` and every child of a node appends a single letter to the string represented by its parent. Logically, Knuth's trie fits this description perfectly. The interesting part about it is how he represents the trie on disk in a very space-efficient manner. Rather than `malloc` a node for every new tree in the trie, Knuth decides to allocate all of the space for the trie up front in four arrays (`LINK`, `SIBLING`, `CH`, and `COUNT`). Trie nodes are associated with indices into these four arrays.

```
...
53: #define TRIE_SIZE 32767
...
60: int LINK[TRIE_SIZE + 1], SIBLING[TRIE_SIZE + 1], COUNT[TRIE_SIZE + 1];
61: char CH[TRIE_SIZE + 1];
...
```

The trie encoding is perhaps most easily described by showing what it looks like after inserting the very first string, `xyz`. On the right we've shown some of the entries in the four arrays, while on the left we've interpreted the important indices into a graph where each index becomes a node, the values in the `LINK` array become solid edges, and the values in the `SIBLING` array become dotted edges.

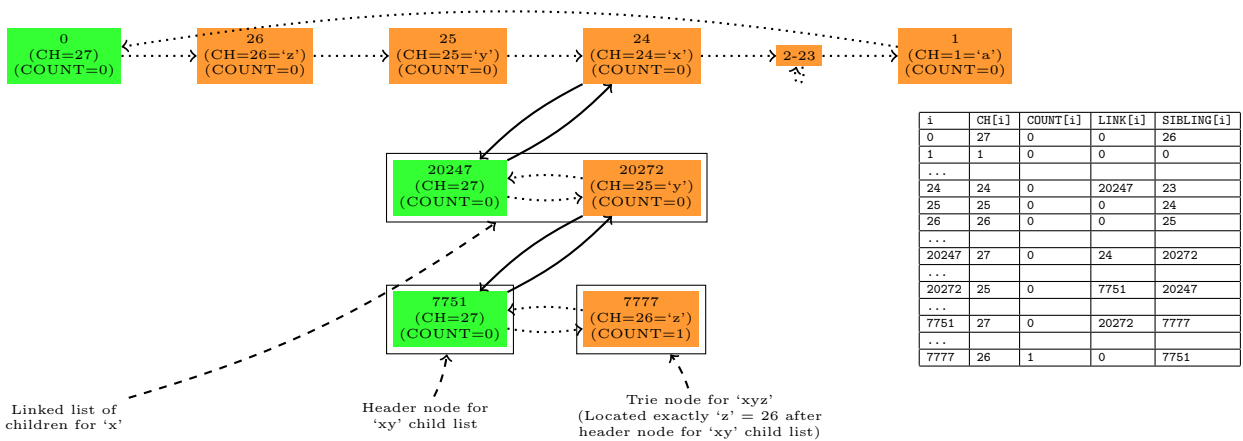

Each node represents an index $i$ where one of `LINK[i]`, `SIBLING[i]`, `CH[i]`, or `COUNT[i]` is nonzero. `LINK[i]` and `SIBLING[i]` are indicated by solid and dotted lines, respectively (zero `LINK[i]` values are ignored). In other words, `1` has a dotted line to `0` because `SIBLING[1] = 0`.

Structurally, there are two distinct types of nodes: trie nodes and header nodes. Orange *trie nodes* represent actual strings (and their prefixes) in the trie. Knuth leaves the root empty string node implicit. He also pre-initializes nodes 1–26 to represent the singleton strings `"a"`, `"b"`, etc., however we've hidden most of these nodes from the logical trie on the left as they've not yet been used. Every trie node can have a circularly linked list containing all of its children. Every such list is represented by a *header node* (green, identified by a `CH` value of 27). Nodes without any children need not have an associated header node (e.g., 7777, representing `"xyz"`, has no children).

## 2.2 Node Overlapping

A naïve implementation would assign each node 27 consecutive slots (one for the header and then one for each of the 26 possible children). But this would waste space in a sparsely branching trie. Instead, when space is running low, Knuth allows the lists to overlap in memory as long as the actually nonempty nodes do not conflict.

It's difficult to see this occur in a small example because Knuth's random number generator ensures the first few lists are well-spaced throughout the array. But if we corrupt the random number generator to always suggest 28 as the initial location guess, we can more readily see its behavior:

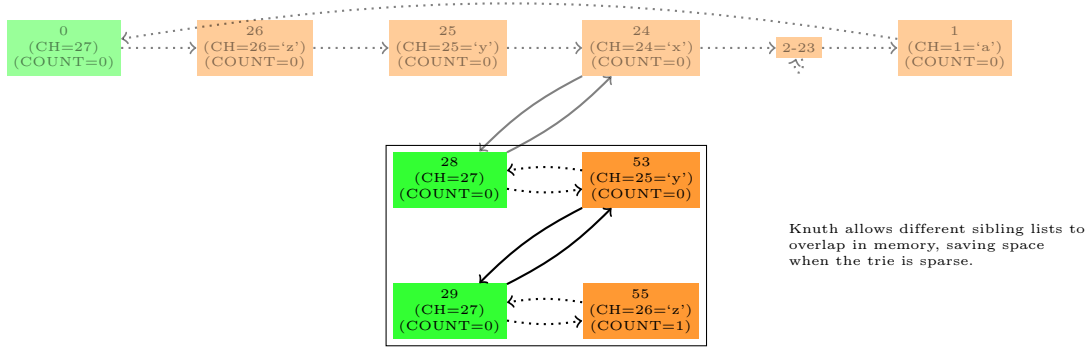

Knuth allows different sibling lists to overlap in memory, saving space when the trie is sparse.

With this worse hash function, we can see that the allocator allows lists to overlap: the header node for `"xy"` (now index 29) is actually placed between the header node for `"x"` (28) and the child node for `"xy"` (53). This means that, if a child node for `"xa"` needs to be created in the future, we'll need to move header node 29 out of the way to make room. While we forced this to happen in this example by manually corrupting the hash function, it can happen even without this corruption (but only once the tables become more full). For now, we'll go back to the uncorrupted random number generator, which produced the trie layout we saw earlier:

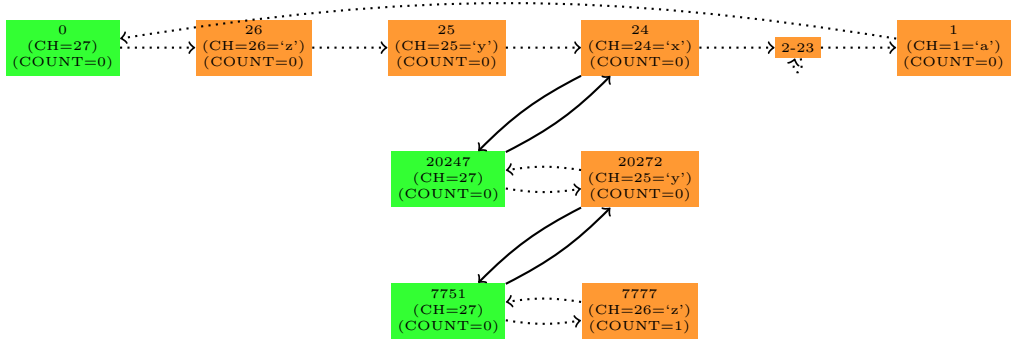

## 3 Inserting the First String

Let's walk through what happens when you insert the first string, `xyz`, into the hash trie. The main function for this is `find_buffer`, which attempts to allocate the node for the one-indexed string stored in the global variable `BUFFER`. Initially, the trie contains a single header node and children for all of the singleton strings `a`, `b`, `...`, `z`.

```
...
79:  int find_buffer() {
80:      int i = 1, p = BUFFER[1], q;
81:      int h, last_h;
82:      while (i < WORD_LENGTH) {
83:          i++;
84:          int c = BUFFER[i];
85:          if (LINK[p] == 0) {
86:              // Insert the firstborn child of p and move to it, or abort_find
...
```

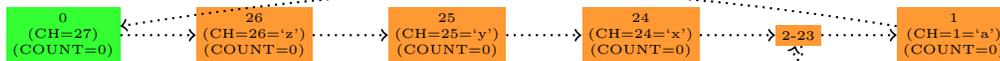

The $p$ variable gets initialized to 24, i.e., the first character in `xyz`, which is conveniently fixed to always be the child node representing the singleton child '`x`'. This node has no children yet, so we start down the 'easy path' of creating a new children list (i.e., header node and one child node). Our first task is to find a location $h$ to place the new header node. Knuth makes an initial guess using a simple random number generator.

```
...
85:          if (LINK[p] == 0) {
86:              // Insert the firstborn child of p and move to it, or abort_find
87:              // Get set for computing header locations
88:              if (X < TRIE_SIZE - 52 - ALPHA) X += ALPHA;
89:              else X = X + ALPHA - TRIE_SIZE + 52;
90:              h = X + 27;
...
```

The global variable $X$ holds the last generated random number and then generates a new one by adding a global constant $\alpha$ with wraparound as necessary. We will start at this initial guess for $h$ and iterate through $h$, $h+1$, $\ldots$, $h + \text{TOLERANCE}$ until we find a location to place both the header node and a child node offset by '`z`' = 24 from the header. If the trie is so full that we cannot find any free space among the first `TOLERANCE` guesses, we give up and return 0.

```
...
91:              if (h <= TRIE_SIZE - 26 - TOLERANCE) last_h = h + TOLERANCE;
92:              else last_h = h + TOLERANCE - TRIE_SIZE + 52;
93:              do {
94:                  // Compute the next trial header location h, or abort_find
95:                  if (h == last_h) return 0;
96:                  if (h == TRIE_SIZE - 26) h = 27;
97:                  else                     h++;
98:              } while (!(CH[h] == EMPTY_SLOT && CH[h + c] == EMPTY_SLOT));
...
```

With Knuth's random number generator, the first location chosen is `h=20247`, which happens to be free. At this point, we need to perform an intricate series of assignments to add the new header and child nodes at location `h=20247`. This is shown step-by-step on the next page.

First, we doubly link the child node for `"x"` (`p = 24`) to the newly created header node for `"x"` (`h = 20247`) and mark it as a header node.

```
...
 99:            LINK[p] = h;
100:            LINK[h] = p;
101:            p = h + c;
102:            CH[h] = HEADER;
103:            CH[p] = c;
104:            SIBLING[h] = p;
105:            SIBLING[p] = h;
106:            COUNT[p] = 0;
107:            LINK[p] = 0;
108:        } else {
...
```

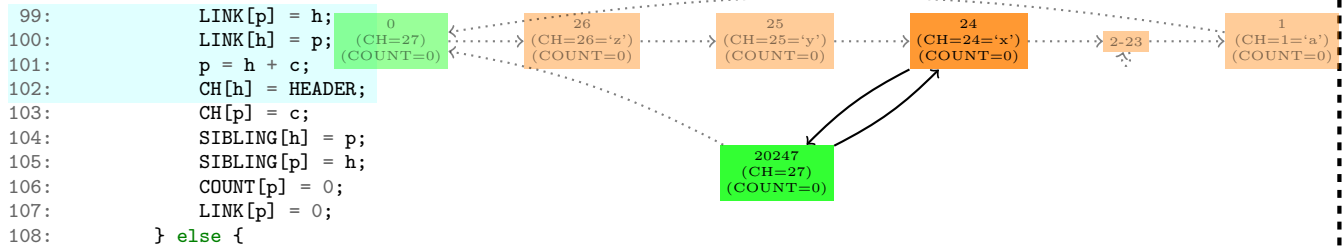

At this point, we've incremented the `p` pointer to where the new child node (for `"xy"`) should go, so we can update its `CH` entry and finish linking it up into a circular list with the new header.

```
...
 99:            LINK[p] = h;
100:            LINK[h] = p;
101:            p = h + c;
102:            CH[h] = HEADER;
103:            CH[p] = c;
104:            SIBLING[h] = p;
105:            SIBLING[p] = h;
106:            COUNT[p] = 0;
107:            LINK[p] = 0;
108:        } else {
...
```

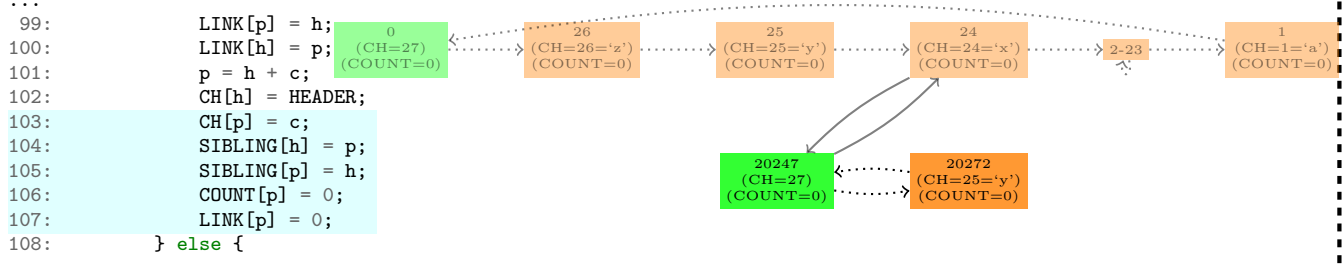

We now have `p` pointing to the child node for `"xy"`, so we repeat the process once more to insert a new child node for `"xyz"`, as desired.

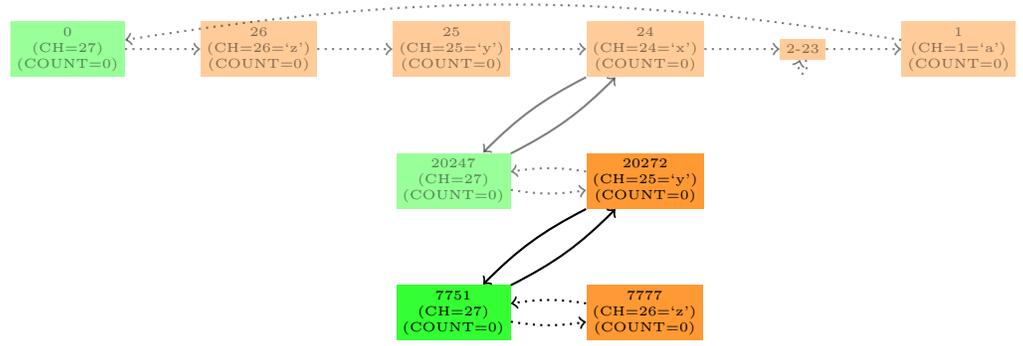

Finally, the trie node representing `"xyz"` (i.e., `p = 7777`) is returned to `main()`, who increments its count value.

```
...
256:        p = find_buffer();
257:        if (p == 0)
258:            WORD_MISSED = 1;
259:        else if (COUNT[p] < MAX_COUNT)
260:            COUNT[p]++;
...
```

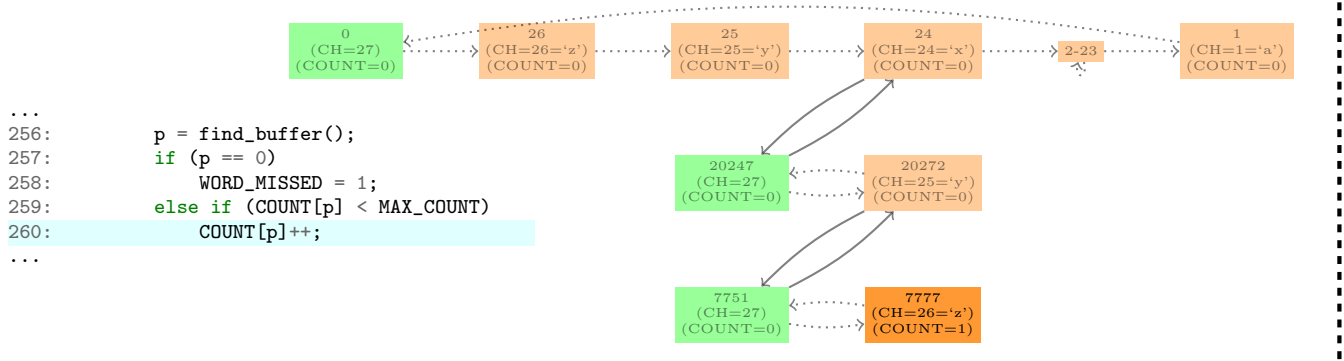

# 4   Handling Conflicts

Let's now consider the most complicated case: inserting into an existing siblings list when the slot you need to insert into is filled by an overlapping siblings list. Recall that, because list headers are inserted at pseudorandom addresses, lists are unlikely to start overlapping until a very large number of nodes have been inserted. **For expository purposes, then, for the rest of this writeup let's once again corrupt the random number generator so it generates overlapping lists as shown below.** Notice again that node 29 (the header node for `"xy"`) is overlapping the list starting at node 28 (the header node for `"x"`).



What will happen if we try to insert the string `"xa"`? First, we'll check if the first character (`'x'`, child node 24) already has any children. It does (we've already inserted `"xyz"`), so we jump to the else condition.

```
...
79:  int find_buffer() {
80:      int i = 1, p = BUFFER[1], q;
81:      int h, last_h;
82:      while (i < WORD_LENGTH) {
83:          i++;
84:          int c = BUFFER[i];
85:          if (LINK[p] == 0) {
...
108:         } else {
109:             q = LINK[p] + c;
110:             if (CH[q] != c) {
111:                 if (CH[q] != EMPTY_SLOT) {
112:                     // Move p's family to a place where child c will fit, or
113:                     // abort_find.
...
```

At this point we have `p=24` and `q=29`, i.e., *q* is pointing to where the second character (`'a'`) *should* be in the children list of `"x"`. Of course, this entry (29) is already in use, by the header for the child list of `"xy"`. It was fine to use that space when we first inserted `"xyz"`, because there was no `'a'` child of `"x"` at the time. But now its presence means we cannot insert an `"xa"` without overwriting the header node for `"xy"`. To resolve this problem, we'll first move the child list of `"x"` to a new location, where we know for sure the offsets we need are free.

At this point, we are trying to insert a new child `"xa"` of the node `"x"`, but the slot that this child would have to be placed in (29) is already occupied by the child list of `"xy"`. We thus need to find a new location (`h`) to move the list currently headed by `p` to. (Once again, Knuth initializes the search for a suitable `h` randomly, but we'll force it to use a crippled random number generator that forces its first guess to be to move the list to $h = 29$.)

```
...
122:                    if (h <= TRIE_SIZE - 26 - TOLERANCE) last_h = h + TOLERANCE;
123:                    else last_h = h + TOLERANCE - TRIE_SIZE + 52;
125:                    // Find a suitable place h to move, or abort_find
126:                    do {
127:                        // Compute the next trial header location h, or abort_find
128:                        if (h == last_h) return 0;
129:                        if (h == TRIE_SIZE - 26) h = 27;
130:                        else                     h++;
131:                        // Find a suitable place h to move, or abort_find
132:                        if (CH[h + c] == EMPTY_SLOT) {
133:                            r = LINK[p];
134:                            delta = h - r;
135:                            while (CH[r + delta] == EMPTY_SLOT && SIBLING[r] != LINK[p])
136:                                r = SIBLING[r];
137:                            if (CH[r + delta] == EMPTY_SLOT)
138:                                slot_found = 1;
139:                        }
140:                    } while (!slot_found);
...
```

The key checks involved are highlighted above. Consider the first iteration of this do-while loop, and suppose by the time we reach line 132 we have $h = 29$ as the first guess for where to move the list to, i.e., shifting the list to the right by one. The code loops over all of the entries in the list and checks whether the location the entry would move to is currently free. Unfortunately, this check fails, because the header itself would then overlap with the existing nonempty entry 29 as shown below, so we have to reject this suggestion and loop again.

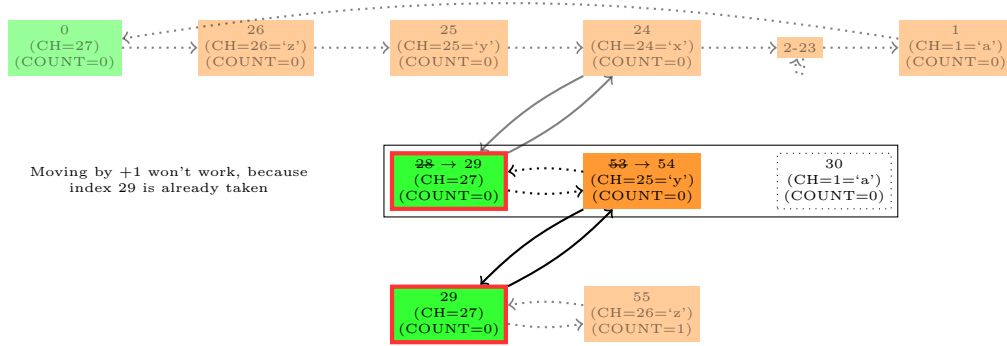

We next consider moving the list by an offset of 2, i.e., placing the header at $h = 30$. Unfortunately, we now find this would cause a conflict with the nonempty entry at index 55 (see below).

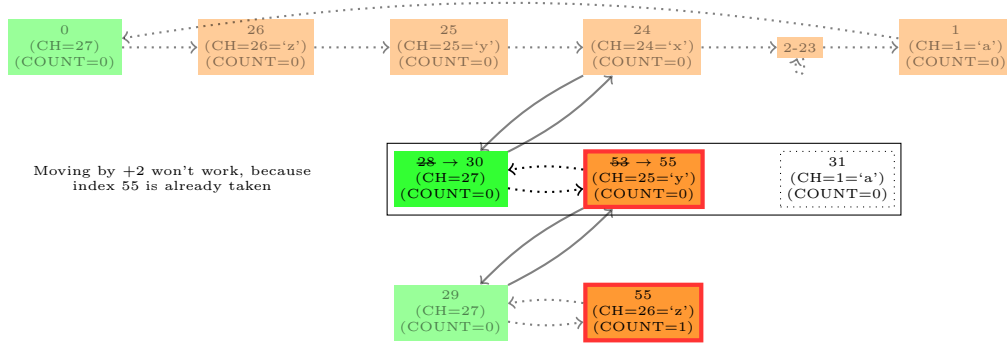

Finally, we consider $h = 31$, i.e., moving the nodes to the right by 3. Thankfully, this shift would cause no conflicts with existing entries, as seen below.

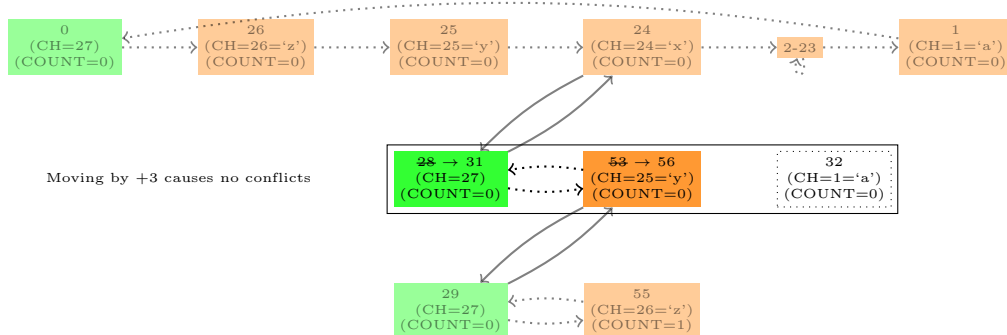

Hence we can set `slot_found=1` and exit the loop successfully.

```
...
137:                          if (CH[r + delta] == EMPTY_SLOT)
138:                              slot_found = 1;
139:                     }
140:                 } while (!slot_found);
142:                 // Move p's family to a place where child c will fit, or
143:                 // abort_find.
144:                 q = h + c;
145:                 r = LINK[p];
146:                 delta = h - r;
147:                 do {
148:                     SIBLING[r + delta] = SIBLING[r] + delta;
149:                     CH[r + delta] = CH[r];
150:                     CH[r] = EMPTY_SLOT;
151:                     COUNT[r + delta] = COUNT[r];
152:                     LINK[r + delta] = LINK[r];
153:                     if (LINK[r])
154:                         LINK[LINK[r]] = r + delta;
155:                     r = SIBLING[r];
156:                 } while (CH[r] != EMPTY_SLOT);
...
```

We're now set up to move each node in the list by a difference of `delta = 3`.

Let's see how the copy happens.

```
...
147:                do {
148:                    SIBLING[r + delta] = SIBLING[r] + delta;
149:                    CH[r + delta] = CH[r];
150:                    CH[r] = EMPTY_SLOT;
151:                    COUNT[r + delta] = COUNT[r];
152:                    LINK[r + delta] = LINK[r];
153:                    if (LINK[r])
154:                        LINK[LINK[r]] = r + delta;
155:                    r = SIBLING[r];
156:                } while (CH[r] != EMPTY_SLOT);
...
```

Initially, we have the following:

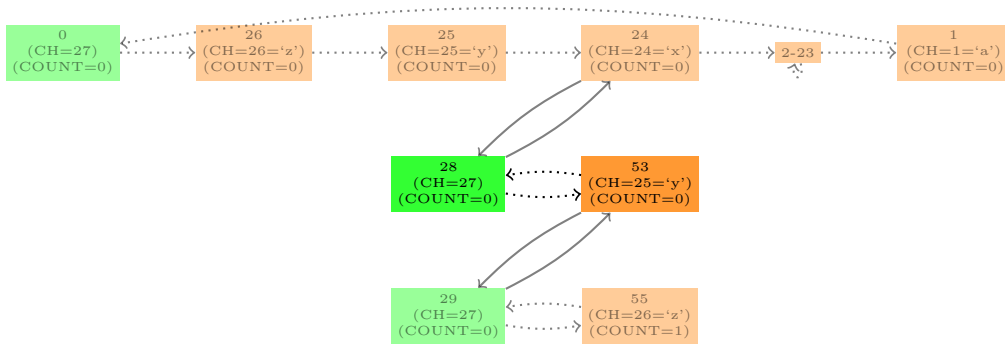

The first iteration copies the header node 28 to 31, updating its sibling node to point to the not-yet-copied-to 56.

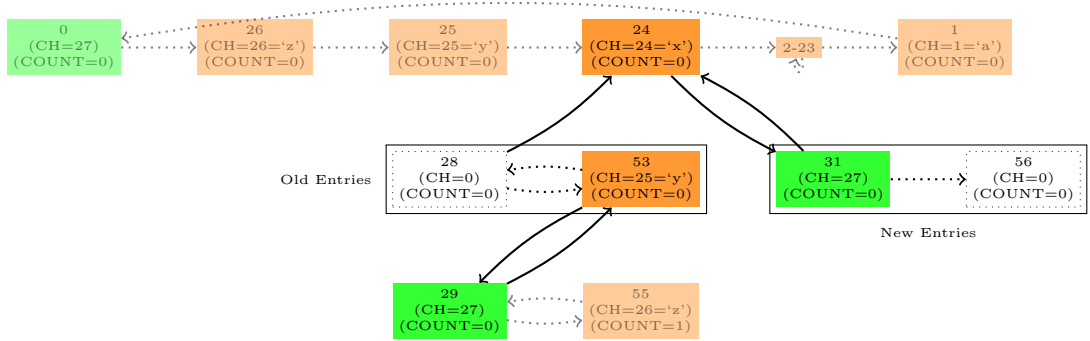

Finally, the second iteration copies 53 to fill slot 56. Once again its `SIBLING` pointer is updated to point to 31 rather than the old header, 28. At this point the old nodes 28 and 53 are no longer reachable from 0, and they both have `CH = 0`, so we can ignore them in our drawing.

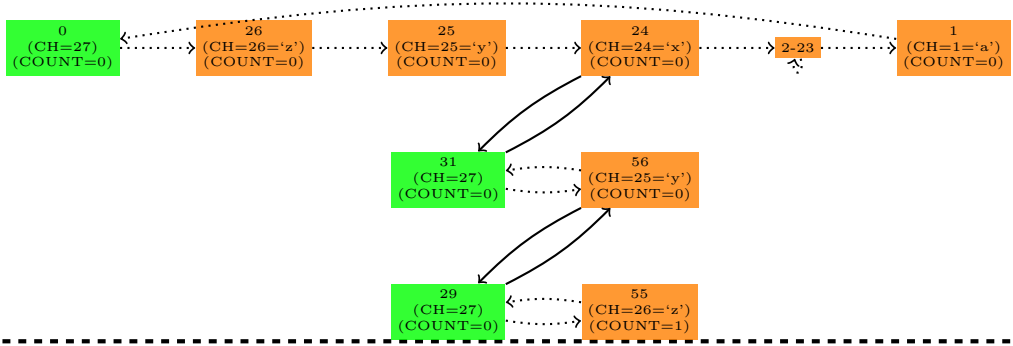

Now that we've moved the list around, we can proceed to add the child node for `"xa"` into the list with header node 31, i.e., at slot 32 which we now see is empty. In comparison to the movement code, the routine to actually insert the new 'a' child into the list headed by index 31 is quite simple. We first find where in the siblings list `"xa"` should be placed in order to maintain the descending order of siblings, then we insert it.

```
...
159:            int h = LINK[p];
160:            while (SIBLING[h] > q) {
161:                h = SIBLING[h];
162:            }
163:            SIBLING[q] = SIBLING[h];
164:            SIBLING[h] = q;
165:            CH[q] = c;
166:            COUNT[q] = 0;
167:            LINK[q] = 0;
168:        }
169:        p = q;
...
```

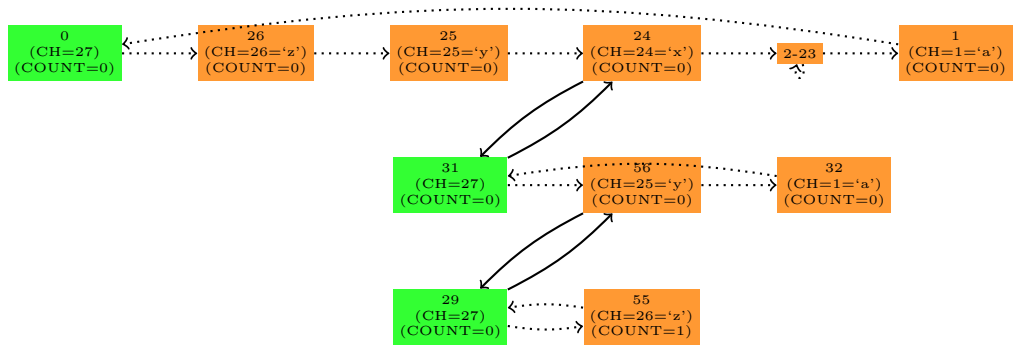

At this point, the index p=32 of the newly added node is returned to `main()`, which increments its count value.

```
...
256:        p = find_buffer();
257:        if (p == 0)
258:            WORD_MISSED = 1;
259:        else if (COUNT[p] < MAX_COUNT)
260:            COUNT[p]++;
...
```

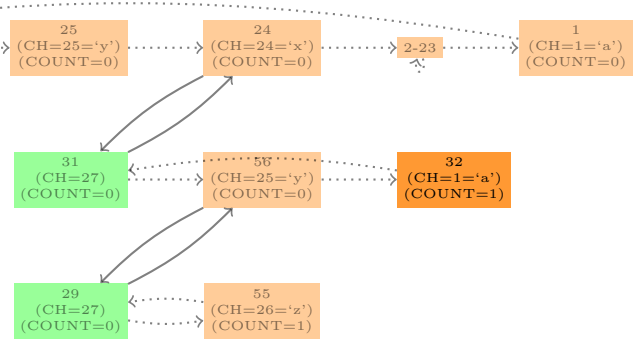



\clearpage

\noindent
We used this shallow-and-wide approach to trace the MiniSAT SAT
solver~\cite{minisat}.
MiniSAT contains nontrivial data structures for variable decision heuristics
and boolean propagation.
Our literate trace covers its main features (variable decisions, boolean
propagation, and clause learning) while glossing over low-level data structure
implementation details.
Glossing over low-level details can frustrate readers, who need to understand
the effect of the skipped-over code before understanding the higher-level
design.
Literate tracing ameliorates this problem: even though some low-level code is
skipped over, \emph{the effects of that code are visible in the concrete state
visualization}.
For example, when a variable is popped from the heap, the
reader can immediately see how the heap changes and that the variable that was
the root of the heap is now the next variable to be guessed.

\subsubsection{Algorithms in the Wild: Explaining a Module}
In a classroom setting, students often want to see that the concepts they are
learning have meaningful applications.
Literate traces can make this connection for students, allowing them to
appreciate the use of a concept in the context of a production system.
Here the goal is not necessarily for the student to come away with a detailed
understanding of the system in question, or be able to make any meaningful
changes to it.
Rather, the goal is for them to see a connection between what they are learning
in lecture and a real system, and give them some jumping-off points (a
callstack or a few key functions) for them to explore more.

In the context of an undergraduate algorithms course we have written literate
traces demonstrating both how GCC~\cite{gcc} uses hash tables and how the
Linux~\citep{linux} kernel uses red-black trees in its scheduler.
Beyond showing students how the algorithms they learn in class can be used in
real software, the traces also proved to be a convenient place to discuss many
different implementation choices (balancing strategies, hash function choice,
intrusive data structures).

In an algorithms class we focus more on the data structure than the application
(scheduling);
a trace written to be shared with an operating systems class would instead
gloss over the red-black insertion code and focus on visualizing the runqueue
itself.
In this way, literate traces have the freedom to focus on whatever is most
important in the given context.

\section{The \trex{} Tool}
\label{sec:Tool}
We have built a tool, \trex{}, that makes writing literate traces easier.
\trex{} extends an underlying documentation language (currently, HTML and
\LaTeX{} are supported) with new commands to interact with a GDB~\citep{gdb}
session and output representations of a running program's state.
\trex{} documents thus involve a combination of three distinct languages:
\begin{enumerate}
    \item The \emph{documentation language} (HTML or \LaTeX{}) where the prose
        and structure of the document are written;
    \item The \emph{program language} (anything supported by GDB) that the
        program being traced is written in; and
    \item The \emph{visualization language} (Python) where the author writes code to
        translate program states into visualizations used by the documentation
        language.
\end{enumerate}

\begin{figure}
    \begin{minipage}{0.45\textwidth}
        \inputminted{}{code_snippets/basic_tex.tex}
    \end{minipage}
    \caption{Literate tracing with the \trex{} \LaTeX{} package}
    \label{fig:BasicTReX}
\end{figure}

\begin{figure}
    \begin{minipage}{0.45\textwidth}
        \inputminted{}{code_snippets/basic_html.html}
    \end{minipage}
    \caption{Literate tracing with the \trex{} HTML preprocessor}
    \label{fig:BasicHTMLTReX}
\end{figure}

\subsection{Basic Usage}
Figures~\ref{fig:BasicTReX}~and~\ref{fig:BasicHTMLTReX} show how to create
basic \trex{} documents using both the \LaTeX{} package and the HTML
preprocessor modes supported by \trex{}.
By default, \trex{} makes available commands like \texttt{setExecutable} (tell
GDB which program to trace) and \texttt{runUntil} (use GDB to run the
program until a given line is reached).
\trex{} packages containing other commands can be imported using
\texttt{trexInitialize} --- in this example, we import the
\texttt{GDBEval} module from the \texttt{built\_in} \trex{} package, which
gives us access to the \texttt{gdbEvalInt} command that reads the value of an
integer expression from the program.
We provide built-in commands for common visualizations:
        \texttt{printCode} pretty-prints snippets of the program's source
        code,
        \texttt{printCallStack} prints a table showing the current call stack
        of the process,
        \texttt{printExpressionTable} prints a table containing the values of
        the specified expressions (most similar to a traditional debugger
        view), etc.
%
One major design goal was staged complexity.
Enough commands are built-in to make it easy for an author familiar with the
documentation language to begin writing a first draft of their literate trace,
while for more advanced visualizations and fine-grained tuning, the author can
define their own program-specific \trex{} modules as described
in~\pref{sec:CustomModules}.

\subsection{\LaTeX{} Package}
The \trex{} \LaTeX{} package augments \LaTeX{} with the ability to set an
executable to be debugged, set breakpoints, and read the state of the executing
program.
Additional plugins can be written in Python using the GDB API to generate
\LaTeX{} code from the executing program's state.

Using the \LaTeX{} package requires two builds of the document.
The first build outputs a special file \texttt{trexout.aux} that contains a log
of all of the \trex{}-related commands in the user's document.
On the second build, our \LaTeX{} package detects that there is an existing
\texttt{trexout.aux} file and runs \trex{} to process the file.
The \LaTeX{} results produced by \trex{} during this processing are then input
into the relevant parts of the \LaTeX{} file during this second build.
This build process is automated by standard tools like \texttt{latexmk} and is
no more complicated than the normal \BibTeX{} double-build process.

\subsection{HTML Preprocessor}
We also provide a \trex{} frontend for HTML.
It acts as a preprocessor, searching through the HTML for \TeX{}-style commands
and interpreting them as \trex{} commands.

The main benefit of the HTML version is the easy creation of \emph{interactive
single-stepping traces} that the user can scrub through to see in detail the
operation of the program~(\pref{fig:SingleStepper}).
Under the hood, interactive single-stepping traces are implemented by running
the program at build time and recording the state every time it reaches one of
a user-specified set of lines.
This is done entirely at preprocessing time: the reader is not running the
program, instead they are seeing essentially a static slideshow of program
steps.

\begin{figure*}
    \centering
\begin{verbatim}
\singleStepper[until=rbtree_augmented.h:84]{rbtree_augmented.h:63-87,rbtree.c}{
    \printProcTree{node,root,gparent,parent,old,new}
}
\end{verbatim}
    \includegraphics[width=0.9\textwidth]{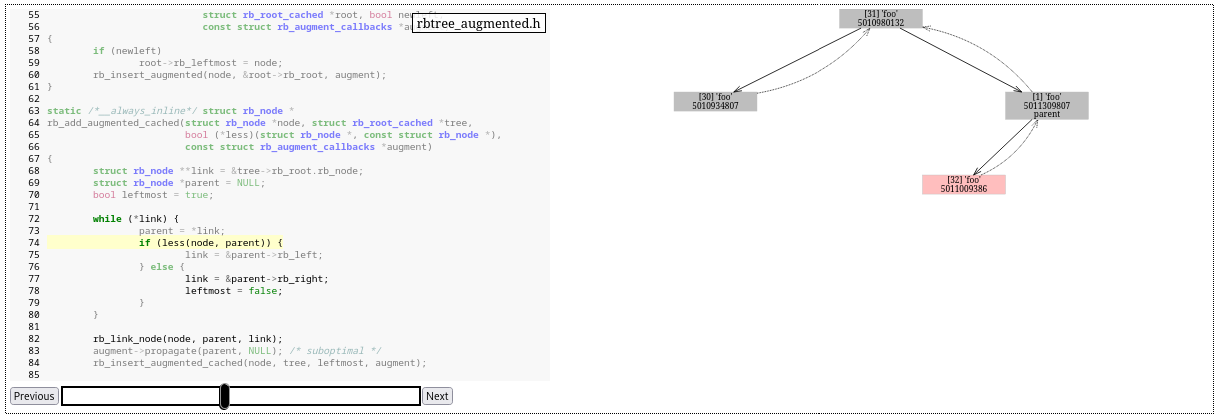}
    \caption{
        This \texttt{singleStepper} command records a frame every time the
        program reaches either (1) any line in the \texttt{rbtree.c} file, or
        (2) lines 63-87 of the \texttt{rbtree\_augmented.h} file.
        For each of those steps, it will output a single frame with the visualization
        computed by the custom \texttt{printProcTree} command.
        %
    }
    \label{fig:SingleStepper}
\end{figure*}

%

\subsection{Writing a \trex{} Module}
\label{sec:CustomModules}
When the built-in \trex{} commands are insufficient, the author can write
custom commands in Python using the GDB Python API~\cite{gdb}.
\trex{} has its own package system shared between the \LaTeX{} and HTML
front-ends.
Every \trex{} \emph{command} belongs to a \emph{module} which lives inside a
\emph{package}.
A package is just a Python file, usually placed in the same directory as the
\LaTeX{} or HTML sources.
A module is a class inside that package inheriting from the \texttt{TReXModule}
base class.
Designating a module method as a command makes it accessible from any \trex{}
document that imports the module.

\begin{figure}
    \inputminted{}{code_snippets/gdb_eval.py}
    \caption{Custom commands can be defined in Python and imported using the
    \texttt{trexInitialize} command.
    Here we show how the built-in \texttt{GDBEval} module can be implemented.
    %
    }
    \label{fig:CustomPackage}
\end{figure}

\pref{fig:CustomPackage} shows a snippet from the included \texttt{built\_in}
package, showing how the \texttt{GDBEval} module we used earlier can be
implemented.
That example is fairly simple, but in general, \trex{} commands can take a long
time to compute (e.g., if they read every entry of a large program array).
To improve build times, if the command does not modify the program state, the
author can add \texttt{"cache": True} to the returned dictionary to tell
\trex{} to cache the result of that command.
The document author can explicitly opt out of using the cache for the next $n$
commands by calling \texttt{\textbackslash{}uncache\{n\}}, or for the rest of
the compilation using \texttt{\textbackslash{}uncache\{inf\}}.

\subsection{Unified Graph Interface}
Many powerful visualizations require the ability to draw graphs.
Unfortunately, there is no good graph-drawing standard shared between both
\LaTeX{} and HTML.
In the \LaTeX{} world the best graph drawing library is TikZ, while in the HTML
world SVG can be used to make graphs.
But the two are not quite comparable: SVG is much lower level than TikZ.
We tried to compile \LaTeX{} documents (including TikZ code) to SVG, but this
slowed down build times.
Graphviz, meanwhile, provides only limited control over graph layout.

We want to give \trex{} authors a single graph interface to target, but still
have it compile down to the best possible format for the actual output target,
and allow authors to fine-tune the graphs in those formats as needed.
To do so, we wrote a Python graphing library that can output to both SVG and
TikZ backends.
The core concepts in the graphing library are \emph{nodes} (boxes with text in
them) and \emph{edges} (between two nodes).
Nodes can be positioned either absolutely (using x--y coordinates), or
relatively (above, below, etc. another node).
This lowering is relatively simple when targeting TikZ, but compiling to SVG
was much more involved: we had to manually draw edge tip arrows, do relative
positioning layouts, and infer the sizes of nodes based on the text.
The result is a relatively simple but powerful graph library usable by both
\trex{} frontends.

\section{Lessons Learned}
\label{sec:Lessons}
This section summarizes lessons we have learned while literate tracing, roughly
structured as a step-by-step guide for writing \trex{} documents.

\subsection{First, Finalize the Code Being Traced}
When preparing the code to be traced, the following checklist is always
important to consider.

\vspace{2mm}
\noindent
\textbf{Does the code take interactive input?}
Change it to read from a file or a hardcoded input so you do not have to
interact with it every time you build the trace.

\vspace{2mm}
\noindent
\textbf{Does the build system pass \texttt{-O0 -ggdb3} (or equivalent) to
the compiler?}
Debug information for optimized code is notoriously poor, and the
\texttt{ggdb3} option enables special GDB-specific features like the use of
macros while tracing.

\vspace{2mm}
\noindent
\textbf{Are there code paths that are hard to reach in a small trace, such
as hash table rebuilds or collision handling code that only happen after a
large number of items are inserted?}
If so, it can be helpful to have two versions of the code: one `real' and one
modified to make it easier to trigger such code (e.g., by decreasing the
default size of the table, or using a random number generator or hash function
that always returns the same number).

\vspace{2mm}
\noindent
\textbf{Does the code have enough comments?}
For \LaTeX{} traces, comments are not very important because you will
likely explain such things in the prose.
On the other hand, HTML traces with single-steppers often leaves the
user on their own while stepping through a chunk of code.
Hence, comments give the user additional local context while single stepping.

\vspace{2mm}
\noindent
\textbf{Does the code put multiple statements on one line?}
Splitting important statements onto their own lines allows breakpoints to be
set on each individually.

\vspace{2mm}
\noindent
\textbf{Does the code declare variables without defining them?}
Making sure variables have default values avoids accidentally visualizing
uninitialized values.

\subsection{Second, Decide What to Trace}
The next useful step is to make an outline of the trace.
For each conceptual step of the code,
decide how much attention you want to pay to it and in how much detail you want
to trace it.
Keep in mind that your reader probably already knows how to accomplish standard
programming tasks, so it makes more sense to spend time and page space focusing
on the nonstandard data structures and algorithms used in the code, or
nontrivial control flow interactions.
This is heavily guided by the imagined reader; see~\pref{sec:Examples} for some
different paradigms.
If the goal is to share a cool data structure trick inside of a scheduler (such
as in the Linux red-black tree trace), there is no need to spend significant
time on unimportant details of, say, how interrupts are handled.

We have found it helpful at this point to build an outline that consists only
of section headers, \texttt{runUntil} commands, and \texttt{printCallStack}
commands.
This ensures that you know what the basic chunks of code you will need to
explain are.
If building an HTML document, it can also be helpful to insert single-steppers
without any state visualizations: this often lets you spend the rest of the
writing process looking only at the draft trace rather than the code itself.

\subsection{Third, Sketch the Visualizations}
You probably want about 2--3 unique classes of visualizations in your \trex{}
document.
Too many and it becomes hard for the user to keep track of what they are seeing
and how it relates to the program internals.
Too few and you risk shoehorning many conceptually different objects into a
single visualization or, worse, boring your reader.

It helps at this stage to sketch out what the visualizations should look like
on a piece of paper.
The more precise your sketches, the less painful it will be to translate them
into a \trex{} module.
For example, if your visualization includes a graph, you should think about how
you want to position the graph nodes, what styles you want for the edges, etc.

Keep in mind that you will probably use the same core visualization in multiple
different parts of the document, so consider what parts of the visualization
are common to all uses and what parts need to be configurable by options
because they change at different points in the document.

\subsection{Fourth, Write Python Code for Visualizations}
At this point, we generally comment out everything in the document outline
except for one location where a custom visualization is needed.
Then implement the visualization in Python and iterate on it until it looks
like the sketch.
Progressively uncomment the rest of the document and add the visualizations
sketched as you go.
Large \trex{} documents can sometimes take a while to build; this strategy lets
you focus on rapidly iterating and refining specific visualization code without
waiting for multi-second builds to complete.

\subsection{Fifth, Add the Prose}
All that is left is to add the surrounding prose explaining to the reader what
is going on.
It is often helpful to add an overview section that walks through the program's
major steps at a high level.
As visualizations become higher and higher level, it also helps to have
sections that explain how those high-level visualizations link up to the
low-level structures in the code (like the hash trie example).

%
%

\section{Limitations and Future Work}
\label{sec:Future}
We now discuss limitations and future work related to \trex{}.

\subsection{Program Evolution and Program Design}

One of the most frustrating tasks when updating the code behind a \trex{}
document is that \trex{} documents refer to line numbers frequently, and those
line numbers change when the code is updated.
We have written a tool that tries to automatically update line numbers after
changes, but perhaps the best ultimate solution would be to integrate
understanding of the \trex{} document into an integrated development
environment (IDE) that updates the line numbers while the user types, or
use Git's native features for tracking line movement throughout the project
history.

This concrete connection to line numbers does, however, have a benefit: unlike
out-of-band design docs, \trex{} documents are \emph{built}, so if they fall
out-of-sync with the main code the build will likely fail.
We could imagine building \trex{} documents as part of CI/CD, forcing
developers to update associated traces when they change the relevant code.

\subsection{Debugger Limitations}
\trex{}'s program state introspection abilities are limited to those provided
the underlying debugger (GDB) and the debug information (DWARF) provided by the
compiler.
It is often impossible to determine whether a particular value is initialized
or not, so illegal values might sometimes appear in a visualization in
misleading ways.
Another issue is when GDB gets line number information wrong, so the effects of
a line might only appear when executing an earlier or later line.
Some source lines might not have corresponding locations in the binary at all,
and the exact sequence of single-step lines might change when a newer compiler
is used.
Thankfully, once the \trex{} document is compiled, it is a static HTML or PDF
document, so these issues are only encountered by the trace authors (not
readers).
Compiler writers are actively working on improving debug information, and many
programs can be built without optimizations to improve debuggability.
We would also be interested in designing a compiler or source-level interpreter
that makes stronger guarantees about the debuggability of its output.


\subsection{Dealing with Nondeterminism}
Programs with nondeterminism are difficult to trace because every time the
trace is rebuilt the resulting document might change.
Sometimes these issues are easy to fix by modifying the application to be
deterministic, e.g., Git can be told to use a fixed timestamp.
In other cases, such as timer interrupts in the Linux kernel, the
nondeterminism is more fundamental to the program operation.
Record-and-replay debugger features can ameliorate some of these issues.
Unfortunately, the recording is often a large binary file that is difficult to
share, and are generally invalidated once the code changes.

Tracing operating systems code introduced its own challenges.
First, we needed to use GDB's ability to connect to a QEMU emulation
session~\cite{qemu}, since the operating system code does not run in user mode.
Second, we had to deal with the fact that operating systems code makes heavy
use of nondeterministic timer interrupts, so every time we built the trace, the
visualizations would look different and perhaps not show what we expect them to
show.
To solve this problem, we were able to use QEMU's record-and-replay feature to
achieve deterministic replays of the code during tracing.

\subsection{Integration with an Interactive Debugger}
Literate tracing with \trex{} often requires the author to write
program-specific state visualizers.
It would be interesting to be able to reuse those visualizations in an
interactive debugging session (i.e., standard GDB).
We envision a future where projects come with \trex{} modules that are used
both for building literate traces and also for adding `superpowers' to GDB when
loaded.
Many projects, e.g., the Linux kernel, already have GDB plugins, but those tend
to be less visual.
It would also be interesting to allow people to record an interactive GDB
session and automatically generate a skeleton literate trace based on that
interactive session.
%


\subsection{Interfaces for Literate Tracing}
With \trex{}, the trace author must program the literate trace in a
documentation language like \LaTeX{} or HTML, and design visualizations in
Python.
It would be interesting to explore an interactive, what-you-see-is-what-you-get
(WYSIWYG) integrated tracing environment that lets users design \trex{} traces
and visualizations in a more graphical manner.

\subsection{Software Archaeology}
There has been a recent push to collect and preserve source code from the early
days of computing history~\citep{swherit,chm}.
While this is a laudable goal, it is not yet so clear what should be done with
this source code.
%
%
We think writing a literate trace of the code could be a useful way to study
and share appreciation of such historical software.

\section{Related Work}
\label{sec:Related}
This section relates literate tracing and the \trex{} tool to other work in
program documentation.

\subsection{Literate Programming}
In literate programming~\citep{litprog,lpcritique}, the program is written
within a prose document that explains its structure.
Literate programming allows nonlinear programming, introducing each line of
code at the pedagogically perfect time for the reader's understanding, which
Knuth argues leads to better code.
Unlike literate tracing, literate programming is a method of
\emph{programming}; it cannot help make existing monolithic codebases (GCC,
Linux, etc.) more understandable.
%
%
%
Furthermore, literate programming tools generally do not deal with runtime
state at all, so they do not provide tooling for making visualizations of
state and they often lack a close connection to any concrete execution of the
program.

\subsection{Software Design Philosophies}
A number of treatises have been written on how to design software to be most
understandable~\cite{aposd,cleancode,oostyle,designpatterns}.
Like literate programming, these techniques do not help one improve the
understandability of software that has already been written, which is the focus
of literate tracing.

\subsection{Notebooks}
Often considered the popular rebirth of literate
programming~\citep{jolly,jupyter2}, we argue that notebooks are actually closer
to an instantiation of literate \emph{tracing}.
Notebooks often only run on a specific concrete dataset, and the notebook
itself executes the code and visualizes results as you read it.
There is a heavy focus on visualizations of the input, intermediate, and output
data.
And much of the code (e.g., the scipy library) is hidden `under the hood' and
not explained by the reader.
\trex{} tries to bring the notebook experience to systems code.

\subsection{In-Line Comments}
Comments are perhaps the most popular way of explaining how source code
works~\citep{commenting,commentsatscale,commentquality}.
They live close to the code in question and are updated
frequently~\citep{updatecomments}, but comments are not perfect.
It is generally much easier for a comment to explain local aspects of the code,
but harder to explain how multiple different modules interact.
And they are usually restricted to plaintext, limiting the ability to draw
useful diagrams.
We see literate tracing as complementary to in-line comments: a literate trace
guides the reader through a system through the help of prose and
visualizations, while comments help the reader who is already reading some
particular part of the code get a sense for what it is aiming to do.

\subsection{Out-of-Band Design Documentation}
Out-of-band documentation and books about system architecture are
popular and effective~\citep{aosa,ulin,gnudecoded,elucidative}.
They avoid many of the limitations of in-code comments, and have been used to
explain the architecture of many large systems.
Without significant author effort, however, out-of-band documentation is
frequently too high-level, not giving the reader enough information about how
and where the high-level architecture is implemented in the actual code.
\trex{} extends existing documentation tools (\LaTeX{} and HTML) with the
ability to refer to and visualize the states of a running instance of the
program.
This makes it easier for documentation writers to visualize program state,
which in turn helps the reader follow what the code is doing at a more concrete
level.
It also ensure fidelity of the documentation to the actual software.
We think that the best documentation strategies mix high-level design and
lower-level tracing, and we believe that \trex{} gives authors the flexibility
to do so.
%


\subsection{Educational Tracers}
Interactive single-steppers are used in educational
contexts~\citep{pytutor,visualgo}.
In~\citep{pytutor} the goal is primarily to explain basic language features
such as pointers, loops, and structs.
Hence, a uniform low-level representation of the program state is often used,
and every single line is steppable.
In~\citep{visualgo} the goal is to explain high-level algorithms, so custom
visualizations are used but the code is generally very short and in an
English-like pseudocode.
\trex{} shows these ideas can scale up to explain components of modern systems.
Instead of focusing on generic visualizations that show all details, we allow
program-specific visualizations and prose that highlight what is most important
about this system in this context.
%

\subsection{Program Comprehension and Visual Debuggers}
Program comprehension and visual debugging research seeks in part to develop
tools that make understanding programs
easier~\cite{importantterms,comprehension1,comprehension2,visualinux,ddd,deet}.
%
%
They are usually only meant to be used by one person, not to generate a
document to be shared with others.
\trex{} is a framework on which system experts can reify their understanding of
a codebase into a single document (the literate trace) that can be shared with
others interested in how the system works.



\section*{Acknowledgements}
This project benefited greatly from insights, conversations, proofreading, and
feedback from the anonymous reviewers, Zachary Yedidia, David K.\ Zhang, Akshay
Srivatsan, Michael Paper, Dawson Engler, and attendees of the Stanford software
lunch.
I am generously funded by grants from the NSF DGE-1656518, Stanford IOG
Research Hub, and Brown Institute.

\bibliographystyle{ACM-Reference-Format}
\bibliography{main}

\end{document}